# Second Quantization Approach to Stochastic Epidemic Models

Leonardo Mondaini[1,2]

[1]Department of Oncology, University of Alberta, Edmonton, Canada

[2]Grupo de Física Teórica e Experimental, Departamento de Ciências Naturais, Universidade Federal do Estado do Rio de Janeiro, Rio de Janeiro, Brazil

Correspondence, Email: mondaini@ualberta.ca, mondaini@unirio.br

**Abstract**
We show how the standard field theoretical language based on creation and annihilation operators may be used for a straightforward derivation of closed master equations describing the population dynamics of multivariate stochastic epidemic models. In order to do that, we introduce an SIR-inspired stochastic model for hepatitis C virus epidemic, from which we obtain the time evolution of the mean number of susceptible, infected, recovered and chronically infected individuals in a population whose total size is allowed to change.

In the pioneering work [1] it was shown that creation and annihilation operators (building blocks of the second quantization method and standard field theoretical language) are not limited to *quantum* systems and, in fact, may be introduced to the description of certain *classical* many-particle systems. Since then, this Fock space formalism (or field theoretical language) for classical systems was complemented by a path integral version and applied to the description of stochastic (birth-death) processes on lattices [2,3].

In this letter we show how this standard field theoretical language based on creation and annihilation operators may be used for a straightforward derivation of closed master equations describing the population dynamics of multivariate stochastic epidemic models. Indeed, our main motivation comes from the fact that, as remarked in [4], for the kinds of model studied in population biology and epidemiology, this field theoretical description is notationally neater and more manageable than standard methods, in often replacing sets of equations with single equations with the same content. As an example of this, we may stress that a single hamiltonian function sums up the system dynamics compactly and may be easily written down from a verbal description of the transitions presented in our model.

Hence, in what follows, we will illustrate the approach above mentioned by introducing an SIR-inspired stochastic model for hepatitis C virus epidemic[1], from which we obtain the time evolution of the mean number of susceptible, infected, recovered and chronically infected individuals in a population whose total size is allowed to change. The number of these individuals will be represented, respectively, by the following random variables: $\mathcal{S}(t)$, $\mathcal{I}(t)$, $\mathcal{R}(t)$ and $\mathcal{C}(t)$.

We will then start by considering a multivariate process $\{(\mathcal{S}(t);\mathcal{I}(t);\mathcal{R}(t);\mathcal{C}(t))\}_{t=0}^{\infty}$ with a joint probability function given by

$$p_{(\{n_i\})}(t) \equiv p_{(n_S,n_I,n_R,n_C)}(t)$$
$$= \text{Prob}\{\mathcal{S}(t) = n_S; \mathcal{I}(t) = n_I; \mathcal{R}(t) = n_R; \mathcal{C}(t) = n_C\}.$$

As previously stated, our aim is to compute time-dependent expectation values of the observables $\mathcal{S}(t)$, $\mathcal{I}(t)$, $\mathcal{R}(t)$ and $\mathcal{C}(t)$, which may be defined in terms of the configuration probability according to

$$\begin{aligned}
\langle \mathcal{S}(t) \rangle &= \sum_{n_S,n_I,n_R,n_C} n_S \, p_{(n_S,n_I,n_R,n_C)}(t), \\
\langle \mathcal{I}(t) \rangle &= \sum_{n_S,n_I,n_R,n_C} n_I \, p_{(n_S,n_I,n_R,n_C)}(t), \\
\langle \mathcal{R}(t) \rangle &= \sum_{n_S,n_I,n_R,n_C} n_R \, p_{(n_S,n_I,n_R,n_C)}(t), \\
\langle \mathcal{C}(t) \rangle &= \sum_{n_S,n_I,n_R,n_C} n_C \, p_{(n_S,n_I,n_R,n_C)}(t).
\end{aligned} \quad (1)$$

The probabilistic state of the system may be represented by the vector

$$\begin{aligned}
\left| \{v_i\} \right\rangle_t &\equiv \left| v_S, v_I, v_R, v_C \right\rangle_t \\
&= \sum_{n_S,n_I,n_R,n_C} p_{(n_S,n_I,n_R,n_C)}(t) \left| n_S, n_I, n_R, n_C \right\rangle \\
&\equiv \sum_{\{n_i\}} p_{(\{n_i\})}(t) \left| \{n_i\} \right\rangle,
\end{aligned} \quad (2)$$

with the normalization condition $\sum_{\{n_i\}} p_{(\{n_i\})}(t) = 1$.

At this point it is worth to note that the fact that the configurations are given entirely in terms of occupation numbers $\{n_i\}$ calls indeed for a representation in terms of second-quantized bosonic operators which then lead us to introduce *creation* and *annihilation* operators for the susceptible, infected, recovered and chronically infected individuals, respectively, $a_i^\dagger$ and $a_i$, $i = \{S, I, R, C\}$, satisfying the following commutation relations

$$\begin{aligned} \left[a_i, a_j^\dagger\right] &= \delta_{ij} \\ \left[a_i, a_j\right] &= \left[a_i^\dagger, a_j^\dagger\right] = 0, \end{aligned} \quad (3)$$

where $\delta_{ij}$ is the Kronecker delta ($\delta_{ij} = 1$ if $i = j$ and $\delta_{ij} = 0$ if $i \neq j$).

---

[1] We will focus on the specific problem of hepatitis C virus epidemic motivated by the recent interest on epidemic models describing this disease, as may be observed in [5]. Note that one of the main differences between our model and the usual SIR epidemic model lies on the fact that, inspired by the observation that approximately 75-85% of people who become infected with hepatitis C virus develop chronic infection [6], we consider the infected state (or compartment) *I* as a kind of metastate from which the individuals evolve to a chronically infected state *C* (in $\sigma \approx 80\%$ of the cases) or a recovered state *R*.

The main idea is that $a_S^\dagger$, $a_I^\dagger$, $a_R^\dagger$ and $a_C^\dagger$ "create", respectively, susceptible, infected, recovered and chronically infected individuals when applied over the reference (*vacuum*) state $|\{0\}\rangle \equiv |0,0,0,0\rangle$, so we can build our space from basis vectors of the form $|\{n_i\}\rangle = \prod_i (a_i^\dagger)^{n_i} |\{0\}\rangle$.

This vacuum state has the following properties: $a_i |\{0\}\rangle = 0$ (whence "annihilation" operators) and $\langle\{0\}|\{0\}\rangle = 1$ (inner product).

Thus, we also have

$$\begin{aligned} a_j^\dagger |\{n_i\}\rangle &= |\ldots, n_{j+1}, \ldots\rangle; \\ a_j |\{n_i\}\rangle &= n_j |\ldots, n_{j-1}, \ldots\rangle, \end{aligned} \quad (4)$$

which imply that $a_j^\dagger a_j |\{n_i\}\rangle = n_j |\{n_i\}\rangle$.

The operators $n_j = a_j^\dagger a_j$ therefore count the number of individuals in a definite state, and are then called *number operators*.

The vector state of our system may be then rewritten in terms of creation and annihilation operators as

$$|\{v_i\}\rangle_t \equiv |v_S, v_I, v_R, v_C\rangle_t$$
$$= \sum_{n_s, n_I, n_R, n_C} p_{(n_s, n_I, n_R, n_C)}(t) (a_S^\dagger)^{n_S} (a_I^\dagger)^{n_I} (a_R^\dagger)^{n_R} (a_C^\dagger)^{n_C} |0,0,0,0\rangle.$$

We may also define a linear operator $\mathcal{H}$ (called *hamiltonian*) which generates the time evolution of our system. This may be easily constructed from the transition rates present in our model according to Table 1 (cf. [4], Table 1).

**Table 1.** Transition rates presented in our model and corresponding terms in the hamiltonian $\mathcal{H}$.

| Transition | Description | Contribution to $\mathcal{H}$ |
|---|---|---|
| $S \xrightarrow{b_S} S + S$ | birth of susceptible individual (rate $b_S$) | $b_S(a_S^\dagger a_S - a_S^\dagger a_S^\dagger a_S)$ |
| $S \xrightarrow{d_S} \varnothing$ | death of susceptible individual (rate $d_S$) | $d_S(a_S^\dagger a_S - a_S)$ |
| $S \xrightarrow{\beta} I$ | infection (rate $\beta$) | $\beta(a_S^\dagger a_S - a_I^\dagger a_S)$ |
| $I \xrightarrow{(1-\sigma)\gamma} R$ | change infected → recovered (rate $(1-\sigma)\gamma$) | $(1-\sigma)\gamma(a_I^\dagger a_I - a_R^\dagger a_I)$ |
| $I \xrightarrow{\sigma\gamma} C$ | change infected → chronically infected (rate $\sigma\gamma$) | $\sigma\gamma(a_I^\dagger a_I - a_C^\dagger a_I)$ |
| $R \xrightarrow{b_R} R + S$ | birth of recovered individual (rate $b_R$) | $b_R(a_R^\dagger a_R - a_R^\dagger a_S^\dagger a_R)$ |
| $R \xrightarrow{d_R} \varnothing$ | death of recovered individual (rate $d_R$) | $d_R(a_R^\dagger a_R - a_R)$ |
| $C \xrightarrow{b_C} C + I$ | birth of chronically infected individual (rate $b_C$) | $b_C(a_C^\dagger a_C - a_C^\dagger a_I^\dagger a_C)$ |
| $C \xrightarrow{d_C} \varnothing$ | death of chronically infected individual (rate $d_C$) | $d_C(a_C^\dagger a_C - a_C)$ |

Upon summing up the terms presented in Table 1, and rearranging them, we may write down our hamiltonian as

$$\mathcal{H} = (b_S + d_S + \beta)n_S + \gamma n_I + (b_R + d_R)n_R + (b_C + d_C)n_C \qquad (5)$$
$$-[b_S n_S a_S^\dagger + d_S a_S + \beta a_I^\dagger a_S + (1-\sigma)\gamma a_R^\dagger a_I + \sigma\gamma a_C^\dagger a_I + b_R n_R a_S^\dagger + d_R a_R + b_C n_C a_I^\dagger + d_C a_C].$$

The equation for the evolution of the probabilities (*master equation* [7]) may then be written as a linear equation in a (imaginary-time) Schrödinger-type form, namely

$$\frac{d}{dt}|\{v_i\}\rangle_t = -\mathcal{H}|\{v_i\}\rangle_t. \qquad (6)$$

Note that by substituting the expressions for the hamiltonian, (5), and the vector state, (2), into the master equation we get after some algebra

$$\begin{aligned}
\frac{d}{dt} p_{(n_S,n_I,n_R,n_C)}(t) &= -[(b_S + d_S + \beta)n_S + \gamma n_I (b_R + d_R)n_R + (b_C + d_C)n_C] p_{(n_S,n_I,n_R,n_C)}(t) \\
&+ b_S(n_S - 1) p_{(n_S-1,n_I,n_R,n_C)}(t) + d_S(n_S + 1) p_{(n_S+1,n_I,n_R,n_C)}(t) + \beta(n_S + 1) p_{(n_S+1,n_I-1,n_R,n_C)}(t) \\
&+ (1-\sigma)\gamma(n_I + 1) p_{(n_S,n_I+1,n_R-1,n_C)}(t) + \sigma\gamma(n_I + 1) p_{(n_S,n_I+1,n_R,n_C-1)}(t) \\
&+ b_R n_R p_{(n_S-1,n_I,n_R,n_C)}(t) + d_R(n_R + 1) p_{(n_S,n_I,n_R+1,n_C)}(t) \\
&+ b_C n_C p_{(n_S,n_I-1,n_R,n_C)}(t) + d_C(n_C + 1) p_{(n_S,n_I,n_R,n_C+1)}(t).
\end{aligned} \qquad (7)$$

The above equation represents the flux of probability between states at rates defined by our model. This kind of equation is also called *forward Kolmogorov differential equation*.

In order to compute the time-dependent expectation values of the observables $\mathcal{S}(t), \mathcal{I}(t), \mathcal{R}(t)$ and $\mathcal{C}(t)$ through the master equation, we need, firstly, to introduce the following *moment generating function (mgf)* [8,9]

$$\begin{aligned}
M(\phi_S, \phi_I, \phi_R, \phi_C; t) &= \langle e^{\phi_S \mathcal{S}(t)} e^{\phi_I \mathcal{I}(t)} e^{\phi_R \mathcal{R}(t)} e^{\phi_C \mathcal{C}(t)} \rangle \\
&= \sum_{\{n_i\}} p_{(\{n_i\})}(t) \prod_i e^{n_i \phi_i}.
\end{aligned} \qquad (8)$$

Multiplying (7) by $\exp(n_S \phi_S + n_I \phi_I + n_R \phi_R + n_C \phi_C)$ and summing on $(n_S, n_I, n_R, n_C)$, leads, after some algebra, to

$$\frac{\partial M}{\partial t} = \left[ b_S\left(e^{\phi_S} - 1\right) + d_S\left(e^{-\phi_S} - 1\right) + \beta\left(e^{-\phi_S + \phi_I} - 1\right) \right] \frac{\partial M}{\partial \phi_S}$$

$$+ \left[ \gamma\left(e^{-\phi_I + \phi_R} - 1\right) + \sigma\gamma(e^{-\phi_I + \phi_C} - e^{-\phi_I + \phi_R}) \right] \frac{\partial M}{\partial \phi_I}$$

$$+ \left[ b_R\left(e^{\phi_S} - 1\right) + d_R\left(e^{-\phi_R} - 1\right) \right] \frac{\partial M}{\partial \phi_R}$$

$$+ \left[ b_C\left(e^{\phi_I} - 1\right) + d_C\left(e^{-\phi_C} - 1\right) \right] \frac{\partial M}{\partial \phi_C}.$$

By differentiating the above equation with respect to $\phi_S, \phi_I, \phi_R, \phi_C$ and evaluating at $\{\phi_i\} = 0$ yields, respectively, the following differential equations for $\langle \mathcal{S}(t) \rangle, \langle \mathcal{I}(t) \rangle, \langle \mathcal{R}(t) \rangle, \langle \mathcal{C}(t) \rangle$:

$$\left[\frac{\partial^2 M}{\partial t \partial \phi_S}\right]_{\{\phi_i\}=0} = \frac{d}{dt}\langle \mathcal{S}(t) \rangle = (b_S - d_S - \beta)\langle \mathcal{S}(t) \rangle + b_R \langle \mathcal{R}(t) \rangle;$$

$$\left[\frac{\partial^2 M}{\partial t \partial \phi_I}\right]_{\{\phi_i\}=0} = \frac{d}{dt}\langle \mathcal{I}(t) \rangle = \beta \langle \mathcal{S}(t) \rangle - \gamma \langle \mathcal{I}(t) \rangle + b_C \langle \mathcal{C}(t) \rangle;$$

$$\left[\frac{\partial^2 M}{\partial t \partial \phi_R}\right]_{\{\phi_i\}=0} = \frac{d}{dt}\langle \mathcal{R}(t) \rangle = (1-\sigma)\gamma \langle \mathcal{I}(t) \rangle - d_R \langle \mathcal{R}(t) \rangle;$$

$$\left[\frac{\partial^2 M}{\partial t \partial \phi_C}\right]_{\{\phi_i\}=0} = \frac{d}{dt}\langle \mathcal{C}(t) \rangle = \sigma\gamma \langle \mathcal{I}(t) \rangle - d_C \langle \mathcal{C}(t) \rangle .$$

The solution of the above set of differential equations has been obtained by using the software *Mathematica* [10].

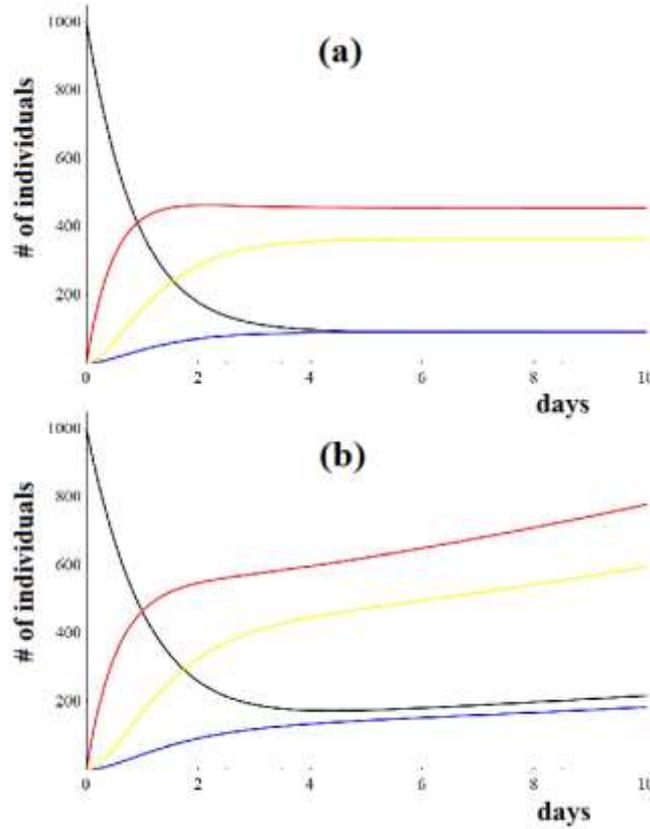

In the previous figures, the solutions for $\langle \mathcal{S}(t) \rangle, \langle \mathcal{I}(t) \rangle, \langle \mathcal{R}(t) \rangle$ and $\langle \mathcal{C}(t) \rangle$ are represented, respectively, by the black, red, blue and yellow lines for $\langle \mathcal{S}(0) \rangle = 1000, \langle \mathcal{I}(0) \rangle = 1, \langle \mathcal{R}(0) \rangle = \langle \mathcal{C}(0) \rangle = 0.$ Note that, in order to obtain these solutions, we have also used the following values for the involved parameters:

(a) $b_S = b_R = b_C = d_S = d_R = d_C = \beta = \gamma = 1.0 \text{ day}^{-1}$ and $\sigma = 0.8$;

(b) $b_S = b_R = b_C = 1.0 \text{ day}^{-1}, d_S = d_R = 0.8 \text{ day}^{-1}, d_C = \beta = \gamma = 1.0 \text{ day}^{-1}$ and $\sigma = 0.8$.

Last but not least, we should stress that the results we obtained allow us to confirm that our model is able to reproduce key features of epidemic processes, including the exponential and asymptotic (plateau) phases. We are presently working on an extension of this work, specifically, trying to generate stochastic sample paths which may be compared to available experimental data.

**Conflict of Interests**
The author declares no conflict of interests regarding the publication of this work.

**Acknowledgments**
This work has been supported by University of Alberta's Li Ka Shing Applied Virology Institute and CNPq, Conselho Nacional de Desenvolvimento Científico e Tecnológico - Brasil.